%% file: main.tex
\documentclass[10pt, conference]{IEEEtran}
\IEEEoverridecommandlockouts
\usepackage{cite}
\usepackage{amsmath,amssymb,amsfonts}
\usepackage{graphicx}
\usepackage{textcomp}
\usepackage{xcolor}
\usepackage{subcaption}  
\usepackage[margin=1in]{geometry}
\usepackage[colorlinks]{hyperref}  
\AtBeginDocument{%
  \hypersetup{pdfborder={0 0 1}, urlcolor=.}  
}
\definecolor[named]{ACMBlue}{cmyk}{1,0.1,0,0.1}
\definecolor[named]{ACMYellow}{cmyk}{0,0.16,1,0}
\definecolor[named]{ACMOrange}{cmyk}{0,0.42,1,0.01}
\definecolor[named]{ACMRed}{cmyk}{0,0.90,0.86,0}
\definecolor[named]{ACMLightBlue}{cmyk}{0.49,0.01,0,0}
\definecolor[named]{ACMGreen}{cmyk}{0.20,0,1,0.19}
\definecolor[named]{ACMPurple}{cmyk}{0.55,1,0,0.15}
\definecolor[named]{ACMDarkBlue}{cmyk}{1,0.58,0,0.21}
\hypersetup{
  colorlinks,
  linkcolor=ACMPurple,
  citecolor=ACMPurple,
  urlcolor=ACMDarkBlue,
  filecolor=ACMDarkBlue
}

\usepackage{inconsolata} 
\usepackage{biolinum} 

\usepackage{microtype}

\def\BibTeX{{\rm B\kern-.05em{\sc i\kern-.025em b}\kern-.08em
    T\kern-.1667em\lower.7ex\hbox{E}\kern-.125emX}}

\input{micro.tex}

\begin{document}

\title{Evaluating LLMs on Sequential API Call Through Automated Test Generation}

\author{
\IEEEauthorblockN{Yuheng Huang\IEEEauthorrefmark{2}, Jiayang Song\IEEEauthorrefmark{3}, Da Song\IEEEauthorrefmark{4}, Zhenlan Ji\IEEEauthorrefmark{5}, Wenhan Wang\IEEEauthorrefmark{6}, Shuai Wang\IEEEauthorrefmark{5}, Lei Ma\IEEEauthorrefmark{2}\IEEEauthorrefmark{1}}
\IEEEauthorblockA{
    \IEEEauthorrefmark{2}
    The University of Tokyo, Tokyo, Japan \\
    \IEEEauthorrefmark{3}
    Macau University of Science and Technology, Macau, China \\
    \IEEEauthorrefmark{4}
    Shandong University, Jinan, Shandong, China \\
    \IEEEauthorrefmark{5}
    Hong Kong University of Science and Technology, Hong Kong, China\\
    \IEEEauthorrefmark{6}
    Institute of Software, Chinese Academy of Sciences, Beijing, China.\\
    \IEEEauthorrefmark{1}
    University of Alberta, Edmonton, AB, Canada\\
yuhenghuang42@g.ecc.u-tokyo.ac.jp, jiayang.song@ieee.org, dasong2296@gmail.com, jiae@cse.ust.hk, \\ wangwenhan@iscas.ac.cn, shuaiw@cse.ust.hk, ma.lei@acm.org}
}

\maketitle

\begin{abstract}
    By integrating tools from external APIs, Large Language Models (LLMs) have expanded their promising capabilities in a diverse spectrum of complex real-world tasks.
    However, testing, evaluation, and analysis of LLM tool use remain in their early stages. Most existing benchmarks rely on manually collected test cases, many of which cannot be automatically checked for rigorous semantic correctness and instead depend on static methods such as string matching. Additionally, these benchmarks often overlook the complex interactions that occur between sequential API calls, which are common in real-world applications.
    To fill the gap, we introduce {\ourmethod}, an automated framework designed to generate diverse coding tasks involving sequential API interactions.
    {\ourmethod} combines state-machine-based API constraint solving and validation, energy-based sampling, and control-flow injection to generate executable programs.
    These programs are then translated into human-like natural language task descriptions through a collaboration of two LLM agents. 
    Utilizing {\ourmethod}, we construct {\ourbench}, a benchmark encompassing 120 manually verified test cases spanning across three representative scenarios: Session Service, Tensor Operation, and ElevenLabs MCP. 
    Experimental results confirm that {\ourmethod} can effectively generate challenging and realistic API-oriented tasks, highlighting areas for improvement in current LLMs incorporating APIs. 
    We make our framework and benchmark publicly available at \href{https://github.com/YuhengHuang42/stateful_bench}{StateEval} to support future research.
\end{abstract}

\begin{IEEEkeywords}
Large Language Models, Code Generation, Tool Use, Sequential Function Call, Benchmark, Fuzzing
\end{IEEEkeywords}

\section{Introduction}
\subsection{Motivation}

The meteoric advancements in Large Language Models (LLMs) have enabled significant performance leaps across domains~\cite{chang2024survey, wang2024survey, naveed2023comprehensive} and have transformed many areas of software engineering~\cite{jiang2024survey, liu2023your, ahmed2022few, ahmed2024automatic, yang2024exploring, pan2024lost}. Building on this progress, stakeholders have found that integrating tool usage with LLM-driven systems can further extend their capabilities~\cite{patil2024gorilla, le2024kat, qin2023toolllm, li2023api}, enabling tasks such as numerical computation~\cite{xu2025llm} and access to up-to-date information~\cite{rag_survey} that are otherwise difficult for LLMs to perform on their own.

\begin{figure}[t]
    \centering
    \includegraphics[width=0.99\linewidth]{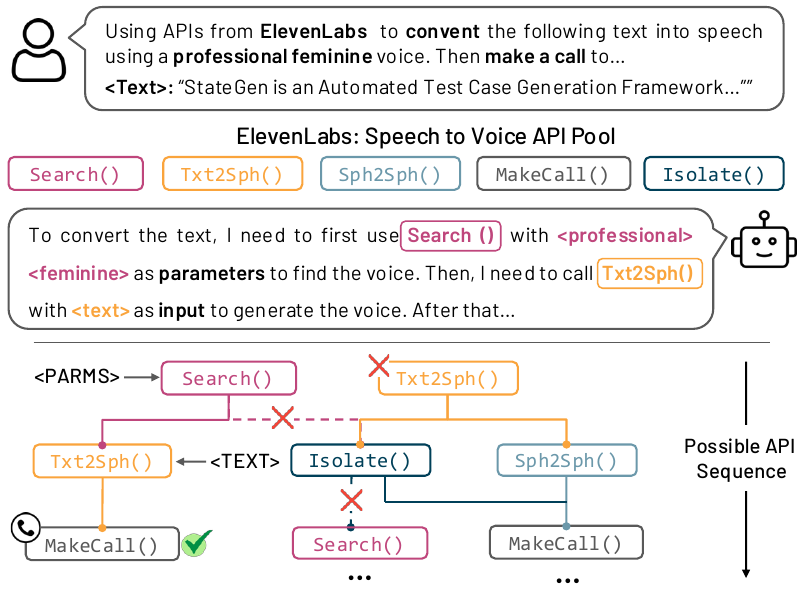}
    \caption{Motivating example of an LLM with sequential API calls.}
    \label{fig:demo}
\end{figure}

Despite the promising progress of this tool integration paradigm, using LLMs to perform designated tasks with external APIs remains a practical yet challenging problem.
In real-world software development, practitioners typically start by analyzing and decomposing the task requirements and then understanding the functionalities and usage of available APIs from their documentation~\cite{zhong2017empirical}.
Eventually, multiple APIs need to be managed in an appropriate sequence, with correctly determined inputs and parameters, to achieve the intended functionalities~\cite{nguyen2014mining, liu2021learning}.
Such a multi-step development pipeline demands advanced abilities in reasoning, management, planning, and tool calling, posing significant challenges for LLMs, which act as the central control unit to coordinate these processes.
As illustrated in Figure~\ref{fig:demo}, converting a given text into speech with a specific style demands the LLM to extract key user requirements (\ie, ``convert to speech'', ``professional feminine voice'') from the input prompt and inspect the available APIs to determine the appropriate ones for execution.
Given the numerous API candidates, the LLM must identify the correct one capable of accomplishing the task and manage the parameters and variables accordingly.
Considering the complexity involved, it is crucial to understand to what extent LLMs can comprehend the functionalities of various APIs and facilitate sequential API calls to fulfill designated objectives. 
Therefore, systematic testing and evaluation of such intricate yet realistic scenarios are one of the foundational components for moving the field forward.



Several recent benchmarks~\cite{liu2023your, chen2021codex, zhuo2025bigcodebench, lai2023ds} have been introduced to systematically evaluate the coding capabilities of LLMs across a variety of contexts. However, most of these efforts primarily focus on general-purpose, small-scale coding tasks, often overlooking scenarios that require reasoning over interdependent API interactions. While some studies~\cite{zan2023private, liu2023codegen4libs, wu2024comprehensive} have begun to incorporate API-centric tasks from specific libraries, their evaluation scope is typically confined to a narrow set of libraries, thereby limiting the generalizability of their findings to broader contexts. 

On the other hand, existing benchmarks tailored for function calling tasks tend to rely on manually curated datasets, which are labor-intensive to construct and difficult to scale to new domains or task types. Moreover, many related works concentrate on addressing daily user requests that involve only a handful of APIs, for example, rather than tackling more complex programming scenarios that demand sophisticated orchestration of multiple APIs. This gap highlights the need for more systematic and scalable benchmarks that can better reflect the challenges encountered in real-world, API-rich programming environments.


Consequently, a key research question then arises:

\textit{How can we design an automated test generation approach that systematically evaluates LLMs' ability in understanding sequential API calls and managing the associated program states?}

\subsection{Challenges}

Addressing this question involves designing a test case\footnote{In this study, we use the term \textit{test case} to refer to a user prompt designed to evaluate an LLM's ability to generate programs involving sequential API calls. We use the term \textit{oracle} to denote the mechanism for assessing the correctness of the generated code.} generation framework that can overcome several key challenges:

\textbf{1. Generating Natural Language Inputs.} Users typically instruct LLMs with informal, natural language. Consequently, the test engine has to be able to automatically generate such prompts to serve as realistic test inputs.

\textbf{2. Managing Complex Program States.} The code corresponding to the realistic user requests involving sequential API calls often has interdependent relationships (\eg, producer-consumer relationship) across different calls. This requires a stateful test engine that can manage program states across these interactions. However, this introduces a tension: the engine must reconcile the informality and ambiguity of user requests with the strict, state-dependent logic of the program.

\textbf{3. Automating Correctness Evaluation.} While existing methods often rely on an LLM-as-judge or simple string matching~\cite{huang-etal-2024-planning-creation}, these approaches can be unreliable for evaluating the correctness of generated programs. We argue that a deterministic, test-oracle-based evaluation is essential for a rigorous and accurate assessment.

\subsection{Contribution}

To address the aforementioned challenges, we propose an automated test generation framework {\ourmethod} and a benchmark {\ourbench} in this study. The {\ourmethod} workflow begins by generating executable programs from existing API documentation. The generation stage, based on a stateful engine, uses API compatibility checking, energy-based API sampling, and control-flow injection to create programs containing sequential API calls, addressing the state management demands in \textit{Challenge 2}. Next, to fulfill the requirements of \textit{Challenge 1}, two LLM agents collaborate to translate these programs into coherent natural language instructions. These instructions then serve as the test inputs for the target LLMs. Finally, oracle construction (\textit{Challenge 3} ) is performed through a replay mechanism: the generated code is executed, and both its final outputs and side effects are recorded as the test oracle. 

Building upon {\ourmethod}, we introduce a new benchmark, {\ourbench}, comprising 120 API-oriented test cases across three representative application scenarios.
In summary, this paper makes the following contributions:













\begin{itemize}[leftmargin=*]
    \setlength\itemsep{0.5mm}

    \item \textbf{Automated Test Generation Framework.} 
    We present {\ourmethod}, a test-generation workflow designed for automatic evaluation of LLMs' capabilities in sequential function calling\footnote{We use API call and function call interchangeably in this paper}. 
    {\ourmethod} integrates a state-machine-based pipeline for program generation and LLM agents for code-to-natural language translation.    

    \item \textbf{Manually Verified Benchmark.} 
    Using this framework, we implement three scenarios: RESTful API calling (\textit{Session Service~\cite{arcuri2019restful}}), tensor manipulation (\textit{Tensor Operation}~\cite{pytorch}), and text-to-speech MCP-based LLM tool calling (\textit{ElevenLabs MCP~\cite{elevenlabs_mcp}}).
    We generate and manually verify 120 test cases, forming a new benchmark, {\ourbench}, to support future evaluation and advancement of LLM-driven API usage.

    \item \textbf{Evaluation and Empirical Study.} 
    We evaluate both {\ourmethod} and {\ourbench} from multiple perspectives using various baselines to validate their effectiveness. 
    Additionally, we report the performance of both open-source and closed-source LLMs on {\ourbench} and provide an initial analysis of their error characteristics.

    \item \textbf{Artifacts.} 
    We publicly release our artifacts, including the implementation of {\ourmethod} (approximately 7,800 lines of code) and our benchmark, {\ourbench} (120 manually verified test cases), via a GitHub repository: \href{https://github.com/YuhengHuang42/stateful_bench}{StateEval}.
    
\end{itemize}

\section{Background}
\label{sec:background}

In this study, we define LLM tool usage as the generation of code involving function calls to accomplish a task. This definition spans two scenarios. The first (Section~\ref {background: llm_SE}) lies in the software engineering context, where LLMs are instructed to produce code with API calls based on engineers’ requirements (\eg, create a new object using a POST request). The second (Section~\ref{background: llm_tool}) concerns enhancing LLM capabilities by integrating diverse external tools as callable functions, enabling LLMs to address general user requests more effectively (\eg, book a flight this Wednesday). The former is more classical and engineering-oriented (\eg, CRUD operations), whereas the latter reflects emerging paradigms and standards (\eg, MCP~\cite{anthropic_mcp}). Despite these differences, we treat both as instances of LLM tool usage and model them within a unified framework.

\subsection{LLM in Software Engineering}
\label{background: llm_SE}
Modern software development is increasingly facilitated by LLMs that can 
(1) synthesize code from natural language and 
(2) leverage external tools and APIs to accomplish complex tasks.
These two capabilities are tightly coupled in practice.
That is, code generation determines how an LLM implements an algorithm or workflow, and tool usage dictates to what extent an LLM can utilize the external functionality and comprehend the hidden sequential workflows.
Effective evaluation, therefore, needs to capture not only function-level correctness but also stateful, sequential behaviors, as well as data and control dependencies when multiple APIs are involved. 

Existing test paradigms for LLM code generation provide useful building blocks~\cite{chen2021codex, liu2023your, austin2021program, jimenez2023swe, zhuo2025bigcodebench} (\eg, unit tests, executable oracles, and string matching), but they often focus on general coding or engineering problems instead of sequential API calls with multi-step interdependencies. 


\subsection{LLM Enhanced with Tool}
\label{background: llm_tool}

By empowering LLMs to use external tools through APIs, we can extend their capabilities beyond static, pre-trained knowledge to perform complex, real-world tasks. This paradigm, explored in comprehensive surveys like ToolLLM~\cite{qin2023toolllm}, shifts the role of an LLM from a text generator to an intelligent agent. Foundational work such as ReAct~\cite{yao2023react} demonstrated that LLMs can effectively combine reasoning traces with task-specific actions (i.e., API calls), while Toolformer~\cite{schick2023toolformer} showed that LLMs can learn to use tools in a self-supervised manner. This enables them to reason, plan, and act upon their environment to solve problems that require up-to-date information or specialized computations.

A critical challenge in LLM tool usage is the necessity for sequential and stateful API interaction. Most meaningful tasks require an LLM to generate not just one, but a sequence of API calls in a logical order~\cite{qin2023toolllm}. 
For example, as illustrated in our motivating example (Figure~\ref{fig:demo}), converting text to speech with a specific voice may first require a \texttt{Search()} API to find a voice ID, which is then passed to a \texttt{Txt2Sph()} API. This process is inherently \textbf{stateful}: the output of one API call becomes a critical piece of the program's state that must be correctly managed and used as an input for a subsequent call. This requirement elevates the task from simple tool selection to a form of program synthesis, where the LLM has to manage data dependencies across multiple steps.

To mitigate errors from incorrect state handling, structured frameworks are emerging. For instance, StateFlow~\cite{wu2024stateflow} models workflows as state machines, but its application has primarily focused on domains like shell (bash) and SQL command execution. This leaves a critical gap for general-purpose API calls, where incorrect state management amplifies the risk of failure, a concern highlighted by safety evaluations like ToolEmu~\cite{ruan2023identifying}. The limitations of existing research and the high stakes involved underscore the urgent need for scalable methods to automatically generate diverse and verifiable test cases from API documentation to systematically evaluate stateful tool use.

\section{Approach}
\label{section:approach}

\subsection{Task Definition of LLM Under Test}
\label{section:approach:problem_def}



We begin by formally defining the problem. An LLM under test is asked to solve a state-dependent code generation task. More specifically, let $S= (s_1, s_2, \ldots, s_n)$ denote the set of all program states, where each $s_j$ corresponds to its respective $j$-th state. A given state may represent either observable program status (\eg, locally defined variables) or external elements (\eg, the data items stored in remote databases). 
Given the natural language instructions specifying the desired tasks for a sequence of API calls, along with documentation that defines the usages of corresponding APIs, the objective for LLMs is to generate code that is syntactically and semantically correct in performing the intended functionalities. 
The generated code should encompass a sequence of $n$ API calls  $O = (o_1, o_2, \ldots, o_n)$, where $o_j$ denotes the API call at the $j$-th timestamp and can have side effects on states in $S$.

\begin{figure*}[t]
    \centering
    \includegraphics[width=\linewidth]{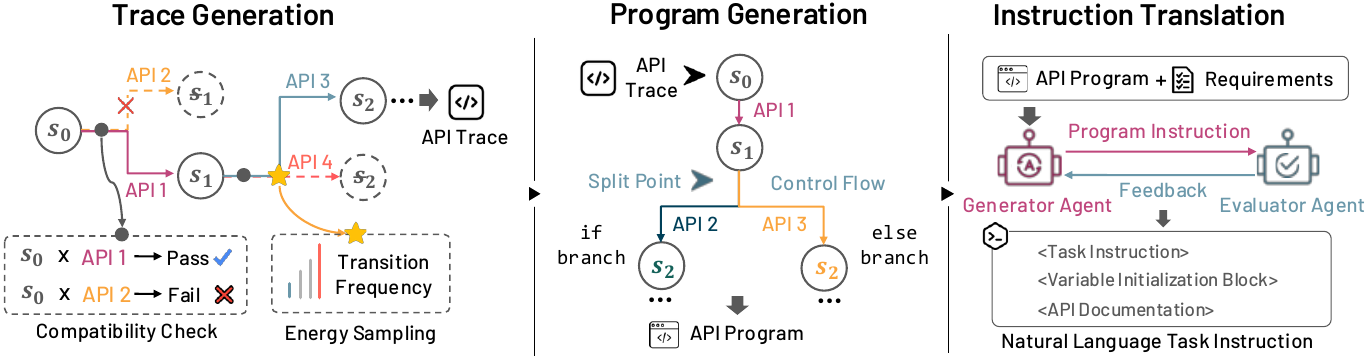}
    \caption{Workflow overview of {\ourmethod}. Trace Generation (Sec~\ref{section:approach:trace}) forms the backbone of {\ourmethod}, producing valid API sequences through state machines while ensuring diversity with energy-based sampling. Program Generation (Sec~\ref{section:approach:program}) then assembles these traces with appropriate initialization and control flow structures. Finally, Instruction Translation (Sec~\ref{section:approach:translation}) employs a multi-agent system to convert the generated programs back into natural language descriptions for evaluation.
    }
    \label{fig:workflow}
\end{figure*}

To evaluate the capabilities of LLMs in addressing such intricate API-oriented coding tasks, we propose {\ourmethod}, an automated test case generation framework for LLMs with sequential API calls. The overall workflow is shown in Fig~\ref{fig:workflow}.
At a high level, {\ourmethod} follows a reverse-generation strategy, beginning with generating valid, executable sequences of API calls of a given length (Section~\ref{section:approach:trace}). 
These traces are then used as a foundation to construct more intricate executable programs by incorporating control flows, thereby enhancing diversity and complexity~(Section~\ref{section:approach:program}).
Next, {\ourmethod} employs an LLM-driven multi-agent translation process (Section~\ref{section:approach:translation}) to transform these programs into human-like natural language instructions. 
To obtain test oracles for automated evaluation, {\ourmethod} executes the generated programs in a local environment and records the state transitions throughout program execution (Section~\ref{section:approach:scenario}). 
These captured transitions serve as ground truth for subsequent evaluation. 

\subsection{Scenario Selection}
\label{section:approach:scenario}

Before presenting our methodology, we highlight a central design decision that shapes our work: instead of collecting existing API traces from real-world systems, as most prior studies do, we build a stateful fuzzing engine that explicitly models how each API behaves \textit{within a scenario}. 
Existing evaluations of LLM tool use tend to emphasize breadth by constructing large API datasets through crawling or repository mining~\cite{basu2025nestful, wu2024comprehensive, saha2024sequential}. Although these datasets cover many tools, they usually treat API calls as isolated or short sequences and often rely on static text matching without execution guarantees. As a result, they overlook the combinatorial complexity of real software, where the challenge lies not only in choosing the right API but also in orchestrating long chains of dependent operations.

To address this limitation, we depart from the ``massive-scale'' strategy and instead adopt a depth-first fuzzing approach. By formally modeling a focused set of domains with state machines, we open a large space of valid execution paths (exponentially growing with respect to sequence length). This enables us to generate long, state-aware API sequences that are structurally sound yet semantically challenging. This dynamic generation paradigm also reduces the risk of data contamination, which is more difficult to avoid when using static benchmarks that an LLM may have already been exposed to during training. 

This design philosophy requires us to carefully select which scenarios to model and fuzz. To reflect the range of challenges seen in real-world function calling, we select three usage scenarios that span distinct stages of technological adoption and specialization. Our selection ranges from mature, general-purpose web standards (such as RESTful APIs) to specialized scientific libraries and emerging LLM-agent-based protocols. This diversity enables us to assess the abilities of LLMs under test across technologies with widely varying levels of abstraction and constraint rigidity, ranging from the explicit specifications of classical web services to the implicit mathematical requirements of deep learning and the schema-driven interfaces of future tool-use systems.

\begin{itemize}[noitemsep, parsep=3pt, partopsep=0pt, leftmargin=*] 
    \item \textbf{Session Service}: This application represents the foundational layer of software infrastructure. We utilize the Session Service from the EvoMaster Benchmark (EMB)~\cite{arcuri2019restful}. As a proxy for a traditional web application, this service manages user sessions via MongoDB using standard RESTful patterns and OpenAPI specifications~\cite{openapi}. By targeting this ubiquitous domain, we validate the LLM's ability to handle stateful resource management within the stable, well-established standards that characterize the majority of existing web APIs. For this scenario, we selected five APIs.

    \item \textbf{PyTorch Tensor Operation}:  This application represents the domain-specific high-performance computing sector. We select PyTorch~\cite{pytorch}, a widely used industry-level deep learning library. Different from general-purpose web services, this domain requires compliance with underlying constraints related to data dimensionality and type handling. LLMs's abilities to track strict mathematical constraints and maintain API compatibility in high-dimensional operations will be evaluated in this application. We select six APIs for this scenario: four related to tensor manipulation and two focused on tensor computation (Linear and Conv2d). Related documentation was sourced directly from the official website.

    \item \textbf{ElevenLabs MCP}: Model Context Protocol (MCP)~\cite{anthropic_mcp} represents the nascent frontier of automated LLLM agent-to-tool interaction. We adopt an implementation from ElevenLabs~\cite{elevenlabs_mcp}. This implementation offers a set of AI tools for speech processing, such as text-to-speech, transcription, and outbound phone calls. As a newly introduced standard designed explicitly for LLM-tool interoperability, MCP differs significantly from legacy APIs in its intention and structure. This scenario evaluates the framework's adaptability to standardized agent interfaces, ensuring that our generation pipeline remains robust when applied to cutting-edge protocols with limited historical training data. From this collection, we choose six APIs.

    
\end{itemize}

These three scenarios span conventional RESTful services, data‑intensive scientific computing, and recent LLM‑driven tool use, covering 17 APIs. Together, they provide diverse, realistic settings to evaluate how LLMs handle API calls and to assess {\ourmethod}. Our choice to orchestrate 5 to 6 distinct APIs per task is also supported by a large empirical study~\cite{zhong2017empirical} of millions of lines of code from popular open‑source projects (e.g., ZooKeeper), which finds that most methods contain short API sequences. Coordinating 5 to 6 APIs can already capture the complexity of most real-world single‑method workflows. Furthermore, {\ourmethod} remains extensible: additional APIs can be included whenever they can be modeled with our state machine.

For clarity, when illustrating our methodology, we will use a simplified \textbf{PyTorch Tensor Operation} example to illustrate the idea.

\subsection{Trace Generation}
\label{section:approach:trace}

Generating valid sequential API calls poses significant technical challenges in this work, as it requires precise modelling of 
program states and meticulous selection of valid API calls at each step. 
Traditional approaches that address this problem have been extensively studied in the context of web service testing, \eg, RESTful API testing~\cite{arcuri2019restful, atlidakis2019restler, viglianisi2020resttestgen, martin2021restest, kim2022automated, liu2022morest, golmohammadi2023testing}. However, these methods are difficult to apply directly, as they focus primarily on CRUD operations, which represent only a subset of our applications. Adapting them for general LLM tool calling would require significant additional effort.

Despite this, the core concept remains unchanged. 
Prior work in RESTful API testing can generally be categorized into top-down~\cite{viglianisi2020resttestgen} and bottom-up~\cite{atlidakis2019restler, liu2022morest} strategies. 
In top-down approaches, the overall API call graph is built in advance, and each API call is instantiated according to the graph. 
Conversely, bottom-up approaches construct the API sequence progressively by adding one call at a time. 
In this work, we adopt a bottom-up strategy, allowing us to employ a ``lazy evaluation'' strategy for determining API inputs and parameters. 
Specifically, we compute only the valid and relevant search space at each step of API selection rather than pre-solving constraints for the entire API call graph, which otherwise would demand substantial computational efforts on optimization to attain practical efficiency~\cite{liu2023nnsmith}.

In particular, {\ourmethod} generates valid sequences of API calls using a component called \textit{TraceGenerator}. To ensure each generated sequence accurately reflects the underlying program logic, \textit{TraceGenerator} maintains a state schema to keep track of $S$, 
which captures all relevant states throughout the trace generation process. The evolution of these states is governed by a series of transitions, where each transition represents an API call.
The transition will have side effects on the state schema, such as modifying, adding, or deleting specific states:

\begin{equation}
[o: \mathbb{S} \times \mathbb{P} \rightarrow \mathbb{S}], \quad \forall o \in \mathbb{O}
\end{equation}
where  $\mathbb{S}$ is the state space and $\mathbb{P}$ is the parameter space of the function call. 
Then, assume that for a specific state variable $s_j$ there are $t$ transitions that have a side effect on it through the entire trace, then we can obtain a series of transformations: 
\begin{equation}
    s^e_{j} = o_j^{t}(o_{j}^{t-1}(\ldots o_{j}^1(s_j^{0})))
\end{equation}
where we denote the initial state as $s_j^{0}$ and refer $s_{j}^{e}$ as the ending state of state variable $s_j$.

\input{algorithm}

For illustration,  we take the \textbf{PyTorch Tensor Operation} as an example. Consider a step in the trace generation process where we have generated the first API call and are about to generate the second (Code Snippet~\ref{illustration:1}).

\begin{codesnippet}[ht]
    \centering
    \caption{First API}
    \label{illustration:1}

    \begin{Verbatim}[
        commandchars=\\\{\},
        numbers=left,
        numbersep=5pt,
        frame=single,  % Adds the box around the code
        rulecolor=\color{black},
        fontsize=\footnotesize
    ]
ut1 = torch.load("user1")
ut2 = torch.load("user2")
\textcolor{commentgreen}{# API 1: concatenate the tensors from}
\textcolor{commentgreen}{# initilized variables ut1 and ut2 }
\textcolor{commentgreen}{# along dimension 0}
r_1 = torch.cat((ut1, ut2), 0)
    \end{Verbatim}
\end{codesnippet}

At this step, the \textit{TraceGenerator} records the state of all intermediate variables, including the shapes and values of tensors \code{ut1}, \code{ut2}, and \code{r\_1}. Furthermore, we update auxiliary information associated with each variable, such as recording that both \code{ut1} and \code{ut2} now participate in a \code{torch.cat} operation. In addition, \code{r\_1} is marked as a newly created variable produced by this function call and has not yet been used as an input to any subsequent operation. Recording such operation history helps the fuzzer track interdependencies between variables, which is essential for selecting the next API call.

By explicitly modelling both the state schema and the transitions, \textit{TraceGenerator} ensures that the generated API sequences are valid and executable. The generation process is outlined in Algorithm \ref{alg:trace_gen}. Initially, \textit{TraceGenerator} starts by creating a collection of random variables to serve as the initial states, which could include user-defined parameters or data items from a remote server (line 2). Note that once the fuzzer finalizes the values of these initialized variables, they are stored persistently on disk and later loaded during evaluation, just as shown in the first two lines of the Snippet~\ref{illustration:1}. This is essential to avoid randomness in the evaluation. 

As the sequence generation progresses, at every step $i$, \textit{TraceGenerator} determines the subsequent API call by iterating through all available APIs and checks whether each one is a feasible next candidate. For example, the \code{conv2d} operation in our \textbf{Pytorch Tensor Operation} example applies a 2D convolution to an input tensor using a weight tensor, with parameters such as \code{stride}, \code{padding}, and \code{dilation}. The \textit{TraceGenerator} checks whether any valid parameter configuration exists that makes this operation legal, given the shapes and types of the existing tensors that can serve as input and weight.  This constraint checking and parameter solving occur on the fly at every step of generation. For \textbf{Pytorch Tensor Operation}, parameter selection is guided by the mathematical rules behind each operation—that is, solving the corresponding equations to determine feasible parameter values. In other scenarios, this process involves checking whether the remote database contains the necessary data items to support a function call, or verifying that the objects currently in scope can serve as valid inputs for the relevant MCP interface.


To further improve the quality and diversity of the generated sequences, we employ energy-based sampling~\cite{bohme2016coverage} strategies when selecting a transition from the candidate pool. This approach is guided by the pair transition; that is, by treating the current transition as a ``consumer,'' we collect all states associated with its input parameters (\eg, ``producers'') and identify the most recent transitions that affected these states. 
This allows us to construct transition pairs in the format \code{(previous transition, current transition)} using states as bridges. Our hypothesis is that increasing the diversity of such transition pairs will enable richer and more complex inter-dependency relationships in generated traces, thereby initiating a more comprehensive evaluation of LLM's ability on API usage. 

Specifically, we employ the following strategies to enhance the diversity of pair transitions. 
If a candidate transition introduces pair transitions that have not been previously observed, it obtains the highest priority upon selection (lines 7–9). 
To further promote diversity, the probability assigned to each transition is inversely proportional to the frequency of its corresponding pair transitions (line 6, with a small $\epsilon$ for numerical stability). 
This approach encourages the exploration of less frequent transition pairs, thereby increasing the overall diversity of the trace generation.
A frequency recorder is used to monitor the appearance of transitions and will be updated upon each selection to reflect the new pair coverage (lines 13-15).

For a more concrete illustration, suppose there are two valid API candidates following Snippet~\ref{illustration:1}, \code{linear} and \code{conv2d}, both of which accept \code{r\_1} as input. Using \code{r\_1} as the linking variable, {\ourmethod} identifies two possible transitions: (\code{cat}, \code{linear}) and (\code{cat}, \code{conv2d}). {\ourmethod} then consults the frequency recorder and observes that (\code{cat}, \code{conv2d}) has been selected less often than (\code{cat}, \code{linear}), and therefore assigns a higher selection probability to the former. If \code{conv2d} is chosen and appended to the trace (as shown in Snippet~\ref{illustration:2}), all relevant state information is updated, and the newly generated variable \code{r\_2} is recorded by StateGen.

\begin{codesnippet}[ht]
    \centering
    \caption{Second API}
    \label{illustration:2}

    \begin{Verbatim}[
        commandchars=\\\{\},
        numbers=left,
        numbersep=5pt,
        frame=single,  % Adds the box around the code
        rulecolor=\color{black},
        fontsize=\footnotesize
    ]
ut1 = torch.load("user1")
ut2 = torch.load("user2")
\textcolor{commentgreen}{# API 1: concatenate the tensors from}
\textcolor{commentgreen}{# initilized variables ut1 and ut2 }
\textcolor{commentgreen}{# along dimension 0}
r_1 = torch.cat((ut1, ut2), 0)
\textcolor{commentgreen}{# API 2: A legal call for conv2d based on}
\textcolor{commentgreen}{# r_1 and ut1}
r_2 = F.conv2d(r_1, ut1, stride=2, 
    padding=0, dilation=1)
    \end{Verbatim}
\end{codesnippet}
\vspace{-5mm}

\subsection{Program Generation}
\label{section:approach:program}

While the trace generated by \textit{TraceGenerator} is already an executable program, we posit that incorporating control flow structures can further reflect the complexity in practical applications and allow for more comprehensive testing of the target LLMs.
The injection of control flows also distinguishes our approach from prior work that relies solely on sequential execution~\cite{li-etal-2023-api, basu2024nestful, patil2024gorilla, wu2024comprehensive, sheng2024measuring}. 

The stateful design schema provided by \textit{TraceGenerator}, makes it straightforward to add branches and other control flow constructs.
One possible strategy involves first selecting a target point from all possible API call sites. 
When trace generation reaches this point, we create a copy of the \textit{TraceGenerator} as the ``else'' \textit{TraceGenerator} with all local variables. 
Then we continue the ``if'' trace generation until it naturally terminates.
The related global data structures from the ``if'' run, such as the frequency recorder, will be passed to ``else'' \textit{TraceGenerator} and complete its generation.  
This process is analogous to spawning a new thread, with both traces proceeding concurrently from the split. 
By repeating this procedure, we can build nested or multiple splits, but in our implementation, we limit a single split to keep the process computationally efficient and easy to manage.

It is worth mentioning that the design of the if-condition at the split point is a non-trivial factor.
To address this issue, {\ourmethod} inspects all local program variables available at the chosen split point in reverse order of creation (\eg, select the most recent variables). 
The chosen variable will be put as the left-hand side of the ``if'' condition, with its runtime value serving as a placeholder on the right-hand side. 
When finalizing the program, this placeholder is replaced by a variable, which is declared in an initialization block at the top of the program (alongside any other randomly initialized variables). 
When collecting test oracles at a later stage, we capture an additional program snapshot by changing this condition variable, thereby flipping into the opposite branch of the program (if it is not a dead code block) to increase test coverage. 
Finally, those unreferenced variables (\eg, unused return values from API calls) will be collected into a RESULT variable, which will be checked during the evaluation stage, preventing silent omission of important side‐effects. In summary, the final outcome of {\ourmethod} includes an initialization block, a program with control flows, and a RESULT variable as the output (\ie, take RESULT as the sink variable).

We illustrate this process using the example in Snippet~\ref{illustration:3}. Suppose the condition insertion point is placed immediately after the second API call in Snippet~\ref{illustration:2}. In this case, {\ourmethod} selects \code{r\_2} as the left-hand-side variable of the if statement, since it is the most recently produced variable, and inserts a placeholder (\code{condition\_1}) for the condition, which will be initialized at the start of the program. After this point, two \textit{TraceGenerator} proceed in parallel, each maintaining its own data structures along its respective branch until the final API is generated. The sink variable RESULT then aggregates the outputs from both branches, which are compared against ground truth during evaluation.

\begin{codesnippet}[ht]
    \centering
    \caption{Full Example}
    \label{illustration:3}

    \begin{Verbatim}[
        commandchars=\\\{\},
        numbers=left,
        numbersep=5pt,
        frame=single,  % Adds the box around the code
        rulecolor=\color{black},
        fontsize=\footnotesize
    ]
\textcolor{commentgreen}{# Initialization Area}
ut1 = torch.load("user1")
ut2 = torch.load("user2")
\textcolor{commentgreen}{## placeholder for the if-condition}
condition_1 = 128
\textcolor{commentgreen}{# API 1}
r_1 = torch.cat((ut1, ut2), 0)
\textcolor{commentgreen}{# API 2}
r_2 = F.conv2d(r_1, ut1, stride=2, 
    padding=0, dilation=1)
\textcolor{commentgreen}{# --->If Condition Insertion}  
if r_2.shape[1] == condition_1:
\textcolor{commentgreen}{    # Traces Generated by TraceGenerator 1}
\textcolor{commentgreen}{    # ...}
    RESULT = ...
else:
\textcolor{commentgreen}{    # Traces Generated by TraceGenerator 2}
\textcolor{commentgreen}{    # ...}
    RESULT = ...
    \end{Verbatim}
\end{codesnippet}
\vspace{-5mm}

\subsection{Instruction Translation}
\label{section:approach:translation}

The next step involves translating the generated program into its corresponding format of natural language instruction, which further reflects the LLM's use cases in the real world. 
To enhance the quality and naturalness of the instruction translation, we employ a multi-agent design comprising a generator agent and an evaluator agent.

In the initial generation phase, the generator agent is employed to produce natural language descriptions based on the initialization blocks and the generated program. This translation process is generally step-by-step, but it may sometimes combine multiple simple API calls into a single sentence or use several sentences to describe a complex API call.
Several requirements are applied within the input prompt to acquire high-quality output: 
(1) unambiguous (\ie, ``the description should enable accurate reproduction of the program''), 
(2) natural and human-like (\ie, ``the descriptions should resemble how developers typically interact with LLMs''), and
(3) non-redundant (\ie, ``only essential information should be provided, allowing the LLMs under test to infer the remaining details'').

Once the generator produces candidate responses, an evaluator agent is prompted to determine whether the generated instructions satisfy the requirements mentioned above. 
The evaluator agent is required to give one of the following three decisions: 
(1) output <OK> to indicate the instructions under assessment are valid; 
(2) if flaws are found, a concise yet specific diagnosis with suggestions should be provided for improvement, or 
(3) output <IMPOSSIBLE> if the agent considers the input description invalid, and generating human-like instructions for the given program is infeasible. 
An iterative bilateral negotiation between two agents will proceed until the evaluator agent produces either <OK>, <IMPOSSIBLE>, or the maximum number of negotiation rounds is reached.

\subsection{Oracle Generation}
\label{section:approach:scenario}

While \textit{TraceGenerator} can use its predicted final states to serve as ground truth, some of its variables may remain abstract (\eg, data item IDs assigned by remote databases). 
Furthermore, there could be subtle discrepancies between our stateful machines and real execution environments. 
To deliver a more accurate and comprehensive assessment, we execute the generated program within the local environment, monitor both explicit and implicit variables, and record their values as oracles. The set of oracles varies across different applications. Specifically:

\begin{itemize}[noitemsep, topsep=0pt, parsep=3pt, partopsep=0pt, leftmargin=*]
    \item \textbf{Session Service}: We deploy the backend service locally, allowing full access to backend states. After each program execution, we record all remote data items. During evaluation, we compare these recorded states to assess correctness.
    
    \item \textbf{Tensor Operation}: High-dimensional tensors are initialized at the start (\ie, loading from disk instead of randomly created) to ensure that results from computations (such as Linear and Conv2d operations) and tensor manipulations are deterministic. At the evaluation stage, we verify both the shapes of tensors and the computation results.

    \item \textbf{ElevenLabs MCP}: Since this scenario involves remote API calls, we employ mock testing for better efficiency and applicability. During evaluation, remote API calls are replaced with a local backend process that records state transitions and generates appropriate responses. We modify both the documentation and APIs to accommodate the mock testing accordingly.
\end{itemize}

\section{Evaluation}

\subsection{Experimental Setup}

\textbf{Implementation.} The majority of {\ourmethod} is implemented in Python, with core functionalities comprising approximately 7,800 lines of code. All experiments are conducted on a server equipped with an AMD Ryzen Threadripper PRO 3955WX CPU (3.9GHz), 256GB of RAM, and four NVIDIA RTX A4000 GPUs, each with 16GB of VRAM. For the implementation of the multi-agent systems, we select \textit{GPT-4.1} for both the generator and evaluator agents. The maximum number of negotiation rounds between agents is set to three.

\textbf{Benchmark Construction.} 
By leveraging {\ourmethod}, a new benchmark, {\ourbench}, is constructed to facilitate the evaluation of LLMs on sequential API calls.
We first generate 60 programs for each of the three scenarios described in Section~\ref{section:approach:scenario}. For each program, we set the number of APIs in the trace to five.\footnote{When an if-else branch is present, the total number of API calls in the program (including both if and else branches) may exceed five. 
However, each individual execution trace, whether following the if or else branch, still contains five API calls.
} 
This number corresponds to the total number of APIs in the Session Service and, based on our observations, represents a moderate level of complexity that is neither trivial nor overly challenging for LLMs. Using too many APIs risks out-of-memory errors during evaluation of locally deployed LLMs, while using too few limits the program's potential complexity.
For the insertion of control flow, we randomly select a point in each program to introduce an if-else branch; a \textit{do not insert} option is also provided to accommodate programs with a single, linear trace. 
This approach is designed to maximize the diversity of the generated programs.

To further ensure the quality of the generated test cases, three authors manually reviewed all 180 (program, instruction) pairs and filtered out those deemed inadequate. 
These unqualified cases may include descriptions that were ambiguous, unnatural, or the translation agent inadvertently assigned initialization values within the instructions rather than merely defining them. 
In total, 7 cases from Session Service, 12 cases from Tensor Operation, and 8 cases from ElevenLabs MCP were filtered out. We then randomly sampled 40 cases from each remaining scenario, resulting in a total of 120 test cases for subsequent experiments. 

This study mainly revolves around four research questions.
RQ1 investigates the effectiveness of our generation framework, {\ourmethod}. 
RQ2 compares our benchmark, {\ourbench}, with existing executable benchmarks for code generation. 
RQ3 assesses the performance of both open-source and closed-source LLMs on our benchmark {\ourbench}.
RQ4 presents a quantitative analysis of the symptoms of LLM errors, and RQ5 follows with case studies that illustrate representative root causes behind these errors.

\subsection{RQ1: How effective is the generation framework {\ourmethod}?}

\begin{figure*}[h!]
  \centering
  \resizebox{0.9\textwidth}{!}{
  \begin{subfigure}[b]{0.32\textwidth}
    \includegraphics[width=\textwidth]{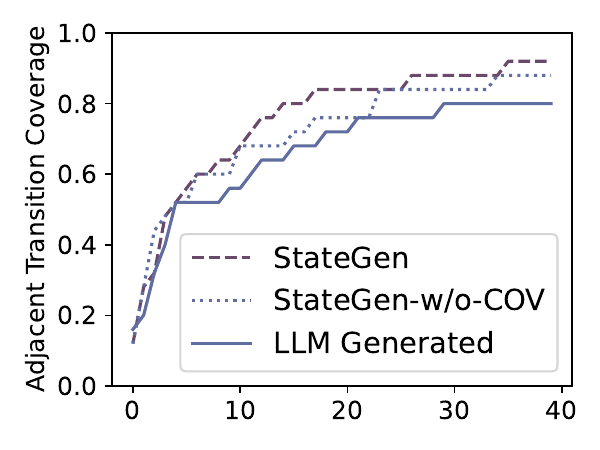}
    \caption{Session Service}
  \end{subfigure}
  \hfill
  \begin{subfigure}[b]{0.32\textwidth}
    \includegraphics[width=\textwidth]{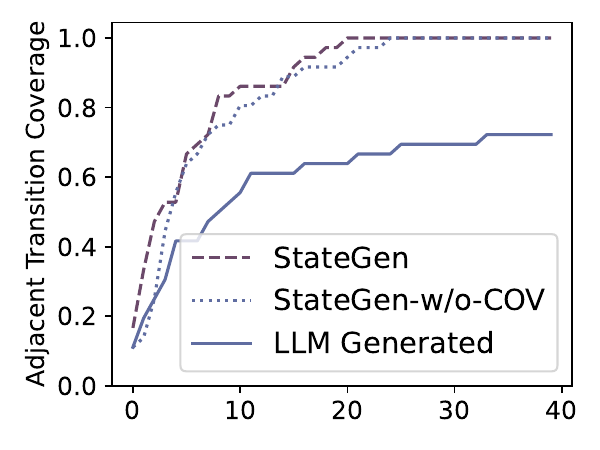}
    \caption{Tensor Operation}
  \end{subfigure}
  \hfill
  \begin{subfigure}[b]{0.32\textwidth}
    \includegraphics[width=\textwidth]{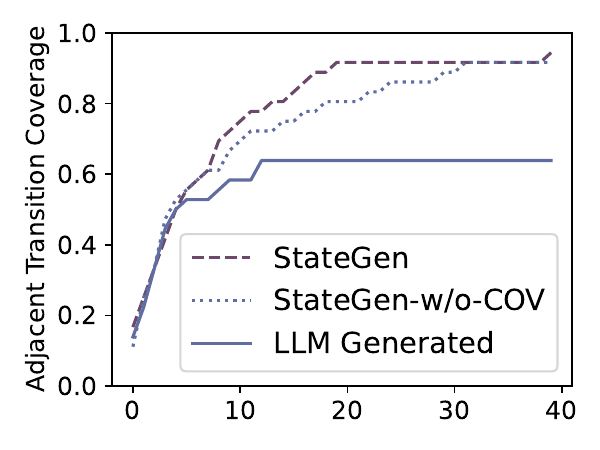}
    \caption{ElevenLabs MCP}
  \end{subfigure}
  \vspace{-10mm}
}
  \caption{Variation in Adjacent Transition Coverage across generated programs for three applications.
  }
  \label{fig:RQ1}
\end{figure*}

In this section, we assess the effectiveness of {\ourmethod} in generating sequences of API calls. As this work represents an early effort to automate the evaluation of LLMs for sequential function call generation, to the best of our knowledge, there are no existing baselines that can be directly adapted for comparison. To address this, we consider two alternative approaches to deliver a comparative evaluation.

For the first approach, we evaluate {\ourmethod} without the pair transition coverage guidance ({\ourmethod}-w/o-COV), as defined in Section~\ref{section:approach:trace}.  
This approach can also serve as an ablation study to assess the effectiveness of pair transition coverage guidance within our framework.
In terms of the second, we implement an LLM-only approach, where LLMs are prompted to generate programs using the documentation for each application. In this setting, we retain the same constraints as in {\ourmethod}, such as restrictions on control flow and the number of API calls. We iteratively query GPT-4.1 until we obtain 40 programs for each application, ensuring the same number of test cases in {\ourbench}. This LLM-only approach is inspired by recent studies highlighting the strong potential of LLMs for automated fuzzing~\cite{xia2024fuzz4all, lemieux2023codamosa}. 
This approach is expected to reveal, without any auxiliary designs, the extent to which individual LLMs can generate diverse test cases for coding tasks with tool usage.

After collecting the programs generated by the LLMs, we observed that many of them required further adjustments and refinements to execute correctly. Common issues included incorrectly defined variables, improper API calls, and the presence of random factors that could lead to flaky tests. Despite this, we can still perform a static analysis to assess the quality of the generated programs.
Specifically, we report the \textit{Adjacent Transition Coverage} (ATC) for such measurement. For a program consisting of $N$ API calls, denoted as $o_1, o_2, \ldots, o_N$, the adjacent transition coverage is calculated as follows:
\begin{equation}
    \text{ATC} = \frac{|\{(o_i, o_{i+1}) \mid 1 \leq i < N\}|}{M^2}
\end{equation}
where $M$ is the total number of unique API calls for a given application. It is important to note that the adjacent transition differs from the pair transition defined in Section~\ref{section:approach:trace}. While pair transition considers APIs bound by states, the adjacent transition only accounts for API calls that occur consecutively (that may not have data dependency). A higher ATC indicates that the generation method produces a broader range of local sequential structures, thereby implying a higher diversity in the generated test cases. This metric is particularly suitable for an initial comparative assessment of program structures generated by different approaches, including those that may not be fully executable or easily analyzable with pairwise transitions. 

We present the results in Figure~\ref{fig:RQ1}. Our results show that {\ourmethod} achieves higher coverage at a faster rate compared to the other two baselines. In particular, {\ourmethod} converges more quickly than the random baseline and is able to discover new adjacent API calls, such as those found in the Session Service. In contrast, the LLM-only approach demonstrates a significantly lower coverage rate than both {\ourmethod} and the random baseline, which indicates a lower diversity in the generated test cases. This finding aligns with recent theoretical work suggesting that LLM-based sampling can be inherently biased~\cite{sivaprasad2025theoryllmsamplingdescriptive}, making it less suitable to serve as an end-to-end fuzzing engine.

\begin{finding}
    \label{finding:1}
    {\ourmethod} demonstrates superior effectiveness in generating sequential API calls, achieving higher coverage and faster convergence compared to other baselines.
\end{finding}

\subsection{RQ2: How does {\ourbench} compare with existing benchmarks?}

To perform the comparison, we collect several widely used, publicly available code generation benchmarks.

\begin{itemize}[noitemsep, topsep=0pt, parsep=3pt, partopsep=0pt, leftmargin=*]

\item  \textbf{HumanEval}~\cite{chen2021codex, liu2023your} is one of the most widely adopted benchmarks, comprising 164 programming problems designed to evaluate LLMs' capabilities in language comprehension and program synthesis.

\item \textbf{DS-1000}~\cite{lai2023ds} includes 1000 data science problems collected from StackOverflow spanning seven Python libraries.

\item \textbf{BigCodeBench}~\cite{zhuo2025bigcodebench} is a recently introduced benchmark that evaluates LLMs' capability of handling diverse function calls and understanding complex instructions.

\item \textbf{Berkeley Function Calling Leaderboard (BFCL)}~\cite{patil2024gorilla} is a popular leaderboard with APIBench, aiming to perform a comprehensive evaluation of the LLM's proficiency in invoking functions and interacting with tools. We select \textit{parallel} and \textit{multiple} executable subsets of BFCL, which have 140 samples.

\end{itemize}

\input{table/dataset_statistics}

We incorporate five metrics to assess the complexity of the collected benchmarks from different perspectives.
In particular, the analysis begins by reporting the token lengths of both the instructions (\eg, input prompts excluding API documentation) and the reference code.
As shown in Table~\ref{tab:dataset}, {\ourbench} exhibits significantly longer input and output lengths compared to the others. 
Although a longer context and generated content do not inherently indicate a more challenging task, they do require LLMs to process more information and reason about the logical and sequential relationships across consecutive instructions. 
This increased complexity is also evident in the average number of function calls, where {\ourbench} nearly doubles the average API call count of the second-highest benchmark, BigCodeBench.

Two metrics, Path Depth~\cite{zhang2011flow} and Binding Count~\cite{beyer2010simple, peitek2021program}, are applied to further characterize the complexity of these benchmarks from both data and structural dependency perspectives. 
Path Depth quantifies the maximum length of a data-dependent chain among API calls, where the output of one call serves as the input to another. A higher Path Depth indicates more deeply nested data dependencies and longer sequential reasoning chains.
Additionally, Binding Count captures the number of shared variables that create data bindings across API calls (\ie, where one call’s output is reused as another’s input). A higher Binding Count suggests a greater density of data interconnections between API calls within a program.
Together, these two metrics reflect the degree of inter-dependency among function calls within a program. 
To compute both metrics, we perform data flow analysis and construct call graphs to identify relevant dependencies. 
Table~\ref{tab:dataset} shows that {\ourbench} exhibits substantially higher values for both Path Depth and Binding Count, indicating that it does not simply stack multiple unrelated API calls. 
Instead, it constructs meaningful dependencies across calls, resulting in more interdependent program structures. It is also worth noting that BFCL yields zero for both metrics, as it typically prompts LLMs to generate API calls that do not share any parameters.

\begin{finding}
    \label{finding:2}
    High-level statistics suggest that {\ourbench} is well-suited for evaluating LLMs' capacity to understand complex instructions and produce multi-API calls with rich interdependencies.
\end{finding}

\begin{table}[t]
\centering
\renewcommand{\arraystretch}{0.85}
\caption{Open-source and closed-source LLMs used in our evaluation.}
\begin{tabular}{@{}p{0.45\columnwidth} p{0.25\columnwidth} >{\centering\arraybackslash}p{0.2\columnwidth}@{}}
\toprule
\textbf{Model Name} & \textbf{Parameters} & \textbf{Availability} \\
\midrule
    Qwen2.5-Coder~\cite{hui2024qwen2} & 32B & Yes \\
    Llama-4-Scout (MoE)~\cite{meta2025llama4} & 17B (/109B) & Yes \\
    Gemini-2.5-Flash~\cite{deepmind2025gemini25flash} & Unknown & No \\
    GPT-4.1-nano~\cite{openai2025gpt41nano} & Unknown & No \\
    GPT-4.1~\cite{openai2025gpt41} & Unknown & No \\
\bottomrule
\end{tabular}
\label{tab:model_overview}
\end{table}
\vspace{-5pt}

\subsection{RQ3: How well do existing LLMs perform?}

\begin{figure*}[t]
    \centering
    \includegraphics[width=0.9\linewidth]{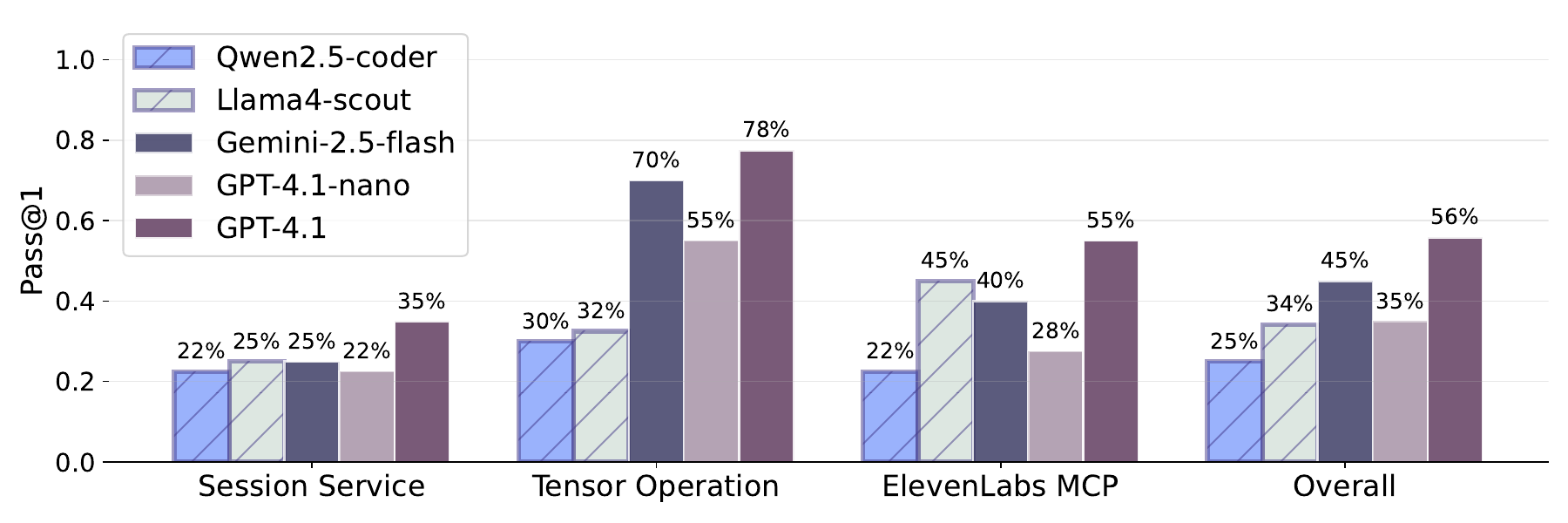}
    \caption{Overall \textit{pass@1} performance of both open-source and closed-source LLMs on {\ourbench}.}
    \label{fig:RQ3_overall}
    \vspace{-3mm}
\end{figure*}

We evaluate in total five LLMs, including two open-source models and three commercial options, using our {\ourbench}. 
The candidate LLMs are shown in Table~\ref{tab:model_overview}. Among these models, Qwen2.5-Coder from the Qwen Team and Llama-4-Scout from Meta are high-performance open-source LLMs that are publicly available and have been widely studied in the literature. For both models, we use the instruction-tuned versions. The remaining models, Gemini from Google and the GPT-4.1 family from OpenAI, are commercial LLMs released after April 2025, representing the most recent advancements.

We use \textit{pass@1}\footnote{Although we use the term \textit{pass@1}, our evaluation includes checks beyond standard assertion-based testing used in other code-generation benchmarks. In addition to verifying functional outputs, we also validate underlying states—such as server-side transitions in MCP workflows or data updates in remote databases. For simplicity, we use \textit{pass@1} to indicate that a generated code snippet satisfies all checks defined in our benchmark.} rates to assess the performance of target models. As shown in Figure~\ref{fig:RQ3_overall}, closed-source models exhibit better capabilities in addressing API-oriented tasks. Specifically, GPT-4.1 achieves the highest \textit{pass@1} rate at 56\%, significantly outperforming the open-source models (25\% for Qwen2.5-Coder and 34\% for Llama-4-Scout). Gemini-2.5-Flash ranks second among all models with a \textit{pass@1} rate of 45\%. In contrast, the cost-efficient GPT-4.1-nano achieves a \textit{pass@1} rate of 35\%, which is slightly lower than that of open-source Llama-4-Scout.

Interestingly, a closer look at the data across tasks shows that the performance of some models varies substantially depending on the task. This variation is particularly pronounced for closed-source LLMs on the Tensor Operation task. For instance, GPT-4.1 achieves a 78\% pass rate on Tensor Operation, which is more than double its performance on Session Service. A similar trend is observed for Gemini-2.5-Flash and GPT-4.1-nano. We hypothesize that this disparity arises because tensor-related operations are fundamental to modern deep learning, resulting in a wealth of publicly available examples and documentation (\eg, higher data availability in the training corpus), especially for frameworks like PyTorch. Consequently, these LLMs are better equipped to handle tensor operations compared to other tasks. 

In contrast, all models perform poorly on Session Service, likely due to the scarcity of online resources and the relatively complex data manipulations required for this task.

\begin{finding}
    \label{finding:3}
    There remains substantial room for improvement for LLMs on {\ourbench}, particularly among open-source models. Although all the generated programs are unlikely to appear in LLMs' training data, greater familiarity with relevant APIs can still significantly improve their performance.
\end{finding}


\subsection{RQ4: What are the symptoms of LLM errors?}

We first conducted a quantitative analysis of the errors made by LLMs on {\ourbench}, categorizing them into three types: 

\begin{figure*}[t]
    \centering
    \includegraphics[width=0.95\linewidth]{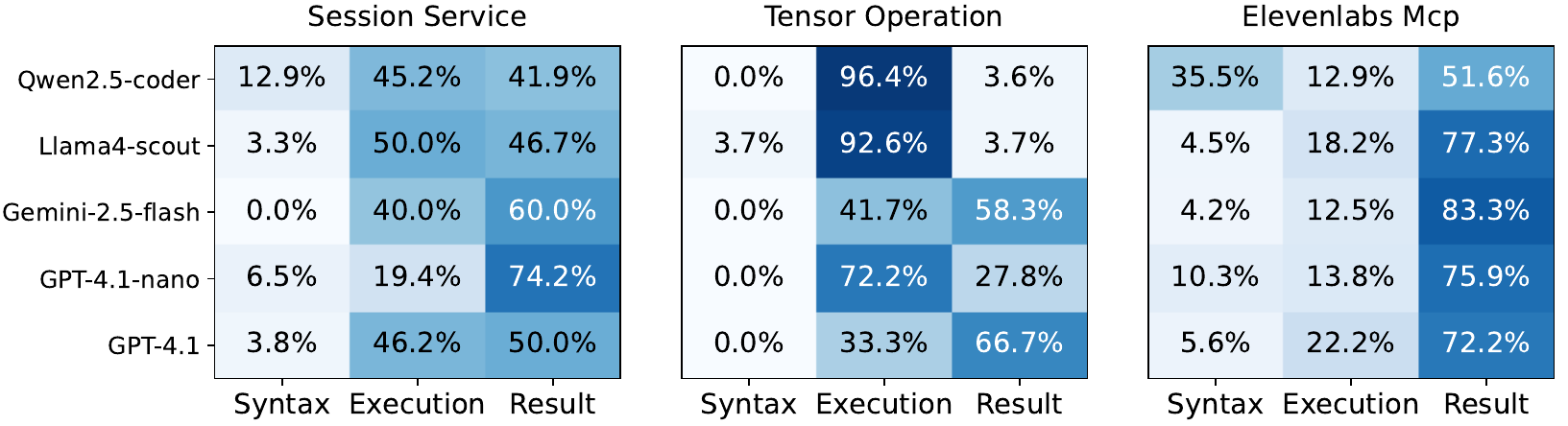}
    \caption{Error type distributions of LLMs. Syntax errors indicate that the generated code contains invalid syntax and cannot be parsed. Execution errors occur when the program starts running but terminates unexpectedly. Result errors refer to cases where the program executes successfully but produces an incorrect final result or side effects.}
    \label{fig:RQ4}
\end{figure*}

(1) \textit{Syntax} errors, where the generated code contains invalid syntax and cannot be parsed; 

(2) \textit{Execution errors}, which occur when a program starts running but terminates unexpectedly, often due to issues with data handling, control flow construction, API selection, or API parameter reasoning; 

(3) \textit{Result errors}, where the program executes without interruption but produces an incorrect final result or side effect, such as failing to update a remote database. The distribution of these error types is shown in Figure~\ref{fig:RQ4}.

In general, execution and result errors are the most prevalent. Execution errors are especially common in Session Service and Tensor Operation tasks. In Session Service, LLMs may attempt to access non-existent data items, resulting in None values and potentially causing downstream errors such as IndexError when operating on empty lists. In Tensor Operation, improper handling of API calls can result in incompatible tensor shapes, causing downstream failures (\eg, RuntimeError encountered for incompatible matrix multiplication in the Linear layer). Conversely, most errors in Elevenlabs MCP are result errors. This is because tool calling in this application is less restrictive regarding input types, but incorrect API usage can still lead to state transition errors, which are only detected during final result checking. For example, the LLMs may fail to place phone calls using the designated voice or may select incorrect parameters during text-to-speech conversion, causing the final speech to not be as expected. On the other hand, syntax errors are relatively uncommon, with Qwen2.5-coder exhibiting them most frequently. This is expected, as Qwen2.5-coder is the smallest model among the weaker open-source ones and can occasionally struggle to adhere to prompt instructions or produce syntax-correct code.

In summary, the test oracles in {\ourbench}, together with the carefully selected scenarios, enable a relatively holistic evaluation of an LLM's ability to generate code involving tool invocations across a broad range of dimensions.

\begin{finding}
    \label{finding:4}
     On {\ourbench}, failures cluster into three types: syntax (unparsable code), execution (crashes from data/control/API or parameter mistakes), and result (runs but yields a wrong outcome). Among these errors, execution and result are dominant while syntax is rare. By spanning diverse, stateful tasks with strict oracles, {\ourbench} can reliably evaluate LLMs across different scenarios.
\end{finding}

\subsection{RQ5: What are the representative root causes of LLM errors?}

We then conducted case studies to analyze the errors made by subject LLMs and identified three primary reasons for their failures: \circled{1} misunderstanding tool documentation, 
\circled{2} not following user instructions accurately, and
\circled{3} mismanaging implicit states across consecutive API calls.

\begin{example*}[htbp]
    \caption{Qwen2.5-coder error on Session Service. Here we show generated code (left) compared with the correct reference code (right).}
    \label{example:1}
    \renewcommand{\arraystretch}{1.1}
    \centering
    \scriptsize
    \setlength\fboxsep{0pt}
    \begin{tabular}{|p{8cm}|p{8cm}|}
    \hline
    {
    \lstset{style=tblstyle, aboveskip=-3pt, belowskip=-8pt, basicstyle=\scriptsize\ttfamily}
    \begin{lstlisting}[language=Python, escapechar=^, numberstyle=\normalsize\color{lightgray}]
    # Generated Code
    def create_session(source, session_type, data):
        ^\colorbox{lightred}{url = f"\{BASE\_URL\}/\{source\}/\{session\_type\}"}^
        response = requests.post(url, json=data)
        response.raise_for_status()
        return response.json()
    # ...
    \end{lstlisting}
    } &
    {
    \lstset{style=tblstyle, aboveskip=-3pt, belowskip=-8pt, basicstyle=\scriptsize\ttfamily}
    \begin{lstlisting}[language=Python, escapechar=!, numberstyle=\tiny\color{lightgray}]
    # Correct Code
    def create_session(source, session_type, data):
        !\colorbox{lightred}{url = f"\{BASE\_URL\}/api/sessions/\{source\}/\{session\_type\}"}!
        response = requests.post(url, json=data)
        response.raise_for_status()
        return response.json()
    # ...
    \end{lstlisting}
    } \\
    \hline
    \end{tabular}
\vspace{-2mm}
\end{example*}

For failure reason \circled{1}, consider a case involving Qwen2.5-coder and the Session Service (Example~\ref{example:1}). The model was tasked with managing remote sessions, and it generated a helper function to handle the POST request for session creation. Although the overall logic was sound, the LLM misread the documentation and constructed an incorrect URL path (\code{/{source/{session\_type}}}), omitting the necessary ``\code{/api/sessions}'' segment. This oversight caused the remote server to return a \code{500 Server Error}. Addressing this issue may require more sophisticated context engineering~\cite{mei2025survey} techniques to help the model make better use of the provided documentation.

For \circled{2} (failed to follow user instructions accurately), we highlight an error made by GPT-4.1-nano on the Elevenlabs MCP task (Example~\ref{example:2}). Although the user explicitly requested the transformation of isolated audio input into speech with a new voice type (that needs to leverage the \code{speech\_to\_speech} API), GPT-4.1-nano incorrectly invoked the \code{text\_to\_speech} tool with text input instead of audio. As a result, the model failed to accurately follow the user's instructions, and the final phone call was incorrectly made. This limitation highlights the importance of further enhancing models’ ability to follow instructions and accurately interpret user requirements.

\begin{example*}[htbp]
    \caption{GPT-4.1-nano error on Elevenlabs MCP. Here we show generated code (left) compared with the correct reference code (right).}
    \label{example:2}
    \renewcommand{\arraystretch}{1.1}
    \centering
    \scriptsize
    \setlength\fboxsep{0pt}
    \begin{tabular}{|p{8cm}|p{8cm}|}
    \hline
    {
    \lstset{style=tblstyle, aboveskip=-3pt, belowskip=-8pt, basicstyle=\scriptsize\ttfamily}
    \begin{lstlisting}[language=Python, escapechar=!, numberstyle=\tiny\color{lightgray}]
    # Generated Code
    search_results = search_voice_library(
        search_category=user_variable_0.get('search_category'),
        search_gender=user_variable_0.get('search_gender')
    )
    if search_results:
        first_voice = search_results[0]
        voice_name = first_voice.get('voice_name')
        isolated_speech = isolate_audio(input_speech=user_variable_0_load)
        # ... Omit for simplicity
        # Instruction: Transform the isolated audio (from
        # user_variable_0_load) into speech
        !\colorbox{lightred}{transformed\_speech = text\_to\_speech(}!
            !\colorbox{lightred}{text=user\_variable\_0\_load,}!
            voice_name=voice_name
        )
        # ...
    \end{lstlisting}
    } &
    {
    \lstset{style=tblstyle, aboveskip=-3pt, belowskip=-8pt, basicstyle=\scriptsize\ttfamily}
    \begin{lstlisting}[language=Python, escapechar=!, numberstyle=\tiny\color{lightgray}]
    # Correct Code
    search_results = search_voice_library(
        search_category=user_variable_0.get('search_category'),
        search_gender=user_variable_0.get('search_gender')
    )
    if search_results:
        first_voice = search_results[0]
        voice_name = first_voice.get('voice_name')
        isolated_speech = isolate_audio(input_speech=user_variable_0_load)
        # ... Omit for simplicity
        # Instruction: Transform the isolated audio (from
        # user_variable_0_load) into speech
        !\colorbox{lightred}{transformed\_speech = speech\_to\_speech(}!
            !\colorbox{lightred}{input\_speech=isolated\_speech,}!
            voice_name=voice_name
        )
        # ...
    \end{lstlisting}
    } \\
    \hline
    \end{tabular}
\vspace{-2mm}
\end{example*}

Finally, Example~\ref{example:3} illustrates the third type of failure (\ie, mismanaging implicit states across consecutive API calls) in the context of tensor operations on Llama4-scout. The model was instructed to perform a sequence of tensor manipulations. However, in the generated code, the model incorrectly inferred the shape of the \code{conv\_result\_2} matrix and inappropriately invoked \code{view} for reshaping. Consequently, this caused a shape mismatch error ("mat1 and mat2 shapes cannot be multiplied") in the linear computation. This mistake caused the code to fail at runtime. Addressing this challenge may require strengthening the model's mathematical reasoning capabilities.

\begin{example*}[htbp]
    \renewcommand{\arraystretch}{1.1}
    \centering
    \scriptsize
    \setlength\fboxsep{0pt}
    \caption{Llama4-scout error on Tensor Operation. Here we show generated code (left) compared with the correct reference code (right).}
    \label{example:3}
    \begin{tabular}{|p{8cm}|p{8cm}|}
    \hline
    {
    \lstset{style=tblstyle, aboveskip=-3pt, belowskip=-8pt, basicstyle=\scriptsize\ttfamily}
    \begin{lstlisting}[language=Python, escapechar=!, numberstyle=\tiny\color{lightgray}]
    # Generated Code
    # ...
    # Output shape: torch.Size([5, 42, 42, 21])
    transposed_tensor = torch.transpose(user_tensor_0, 1, 3)
    if list(user_tensor_0.shape) == user_constant_0:
        # ... Omit for simplicity.
    else:
        # ... Omit for simplicity.
        # Output shape: torch.Size([5, 24, 42, 21])
        conv_result_2 = F.conv2d(transposed_tensor, user_weight_2, stride=1, padding='same', dilation=1)
        # user_weight_3 shape: torch.Size([21, 12])
        !\colorbox{lightred}{linear\_result = F.linear(conv\_result\_2.view(-1, 42), }! !\colorbox{lightred}{user\_weight\_3)}!
        # ...
    \end{lstlisting}
    } &
    {
    \lstset{style=tblstyle, aboveskip=-3pt, belowskip=-8pt, basicstyle=\scriptsize\ttfamily}
    \begin{lstlisting}[language=Python, escapechar=!, numberstyle=\tiny\color{lightgray}]
    # Correct Code
    # ...
    # Output shape: torch.Size([5, 42, 42, 21])
    transposed_tensor = torch.transpose(user_tensor_0, 1, 3)
    if list(user_tensor_0.shape) == user_constant_0:
        # ... Omit for simplicity.
    else:
        # ... Omit for simplicity.
        # Output shape: torch.Size([5, 24, 42, 21])
        conv_result_2 = F.conv2d(transposed_tensor, user_weight_2, stride=1, padding='same', dilation=1)
        # user_weight_3 shape: torch.Size([21, 12])
        !\colorbox{lightred}{linear\_result = F.linear(conv\_result\_2, user\_weight\_3)}!
        # ...
    \end{lstlisting}
    } \\
    \hline
    \end{tabular}
\vspace{-2mm}
\end{example*}

\begin{finding}
    \label{finding:55}
    Our case studies reveal three representative root causes behind LLM failures in generating code with tool invocations. First, models can sometimes misinterpret API documentation, overlooking required conditions. Second, in certain scenarios, they fail to follow users' instructions, leading to incorrect operations. Third, they may struggle with interdependent API calls, producing inconsistent data flows or broken dependencies.
\end{finding}

\section{Discussion}

\subsection{Limitation and Future Work}

(1) \textbf{Engineering effort for incorporating new scenarios}: Our state machine currently requires manual modelling of each API's documentation before it can generate programs for different scenarios. This process can be labor-intensive. Future work could reduce this effort by developing LLM-enabled middleware that automatically parses API documentation and translates it into structured state and transition models. While we acknowledge this limitation, \textit{our framework can still generate an extensive range of test cases within the existing three scenarios}. This diversity stems from our fuzzer-like test generation engine. As the number of APIs in a program sequence increases, the number of diverse programs expands exponentially. Our engine can explore this large space by varying not only the API calls but also the program initial states and the API parameter combinations. This allows our current artifact to be used for extensive testing of subject LLMs.

\noindent (2) \textbf{More in-depth analysis}: Future work can extend our benchmark in three directions: (1) evaluate a broader range of models and prompting techniques, including advanced reasoning strategies (e.g., chain-of-thought); (2) develop automated root cause analysis to detect issues in generated code and enable self-reflection; and (3) improve the correctness and robustness of LLM‑generated sequences of function calls, especially in complex or compositional settings.

\noindent (3) \textbf{Beyond correctness evaluation}: While our work centers on code correctness, the framework's core methodology—combining automata-based stateful generation with LLM-based transformation—is adaptable for evaluating other critical properties. For example, the fuzzing engine can be modified to generate malicious code or programs with vulnerabilities, effectively turning the framework into a tool for security testing. Likewise, it can assess privacy risks by having an LLM inject sensitive information into the generated code to test for potential data leaks.

\subsection{Threat to Validity}
\textbf{External Validity.} A potential threat to external validity arises from the possibility that {\ourmethod}'s effectiveness in generating diverse and challenging test cases may not generalize to different kinds of APIs. To mitigate this, we selected three substantially different domains—RESTful API calls (Session Service), tensor computations (Tensor Operation), and multimodal tool use (ElevenLabs MCP)—to introduce domain-level diversity and improve the generalizability of {\ourmethod}.



\noindent\textbf{Internal Validity.} 
A threat to internal validity stems from potential variability in the natural language instructions generated by LLMs during the construction of benchmark tasks. Inconsistent or ambiguous instructions could confound evaluation results by introducing unintended complexity. To address this, {\ourmethod} incorporates a multi-agent negotiation protocol designed to produce clear and coherent instructions. Furthermore, we manually verified all 120 test cases in StateEval to ensure instruction clarity, semantic accuracy, and task consistency across scenarios.

\noindent\textbf{Construct Validity.} 
A potential threat to construct validity is whether the selected metrics truly capture the intended concepts of benchmark complexity and LLM proficiency. To address this, we evaluate from multiple perspectives. For benchmark complexity, we combine general metrics, such as instruction length and number of functions, with specialized metrics like Path Depth and Binding Count. For LLM proficiency, we complement the \textit{pass@1} with quantitative analyses of observed behaviors and case studies that examine underlying root causes.

\section{Related Work}

In this section, we review related work from three perspectives: (1) benchmarks for evaluating the correctness of generated programs by LLMs, (2) studies assessing LLMs' ability to perform API calls, and (3) RESTful API fuzzing methods that involve sequential trace generation.

\subsection{LLM Code Generation and Benchmark}
\label{related_work: code_llm}
LLMs have led to a significant performance leap in code synthesis, with models such as Gemini 2.5~\cite{comanici2025gemini} and Qwen-Coder~\cite{hui2024qwen2} demonstrating strong performance on programming tasks.
Despite the swift advancements, the evaluation of such coding capabilities remains fragmented across task granularity, complexity, and oracle quality. 
In particular, the evaluation of LLM code generation, at the early stage, focused on function-level code synthesis with executable unit test suites.
HumanEval~\cite{chen2021codex, liu2023your} presented the widely-accepted \textit{pass@k} metric using manually crafted Python problems, offering execution-based oracles for assessing function correctness.
The MBPP~\cite{austin2021program} expanded its range to include approximately 1000 beginner Python tasks, each accompanied by three test cases, serving as a standard entry-level benchmark for evaluating the coding ability of LLMs.

In addition to function-level synthesis, practitioners constructed benchmarks that included libraries, APIs, and more stringent oracles to better reflect the complexity of real-world software development processes.
DS-1000~\cite{lai2023ds} collected 1000 programming problems associated with seven Python libraries (\eg, NumPy and Pandas) from StackOverflow and introduced reliable automatic evaluation by combining test suites and constraints on API usages or keywords. 
Meanwhile, HumanEval-XL~\cite{peng2024humaneval} expanded the benchmark and evaluation to a multilingual context, thereby investigating the cross-lingual generalization of code LLMs.

Building on function-level code synthesis, SWE-bench~\cite{jimenez2023swe} pioneered a repository-level assessment utilizing actual GitHub repositories and issues. 
That is, LLMs are required to create patches for projects that involve multiple files.
Complementarily, {BigCodeBench}~\cite{zhuo2025bigcodebench} focuses on assessing the capabilities of code LLMs by utilizing functions from diverse libraries.

\subsection{Evaluating LLMs with API Call}

Substantial efforts have been made by practitioners to assess the capability of LLMs incorporating APIs under various scenarios~\cite{qin2023toolllm,openaiFC,mistralFC,llamaFC} with natural language instructions. 
Nevertheless, most existing approaches focus on building benchmarks or performing empirical studies by either manually collecting test cases or extracting them from online forums~\cite{li2023api, patil2024gorilla,basu2024nestful,huangmetatool,zhan2024injecagent,shen2024shortcutsbench, zhong2024can}.

Early studies have focused on evaluating LLMs’ ability to select the appropriate tool and invoke it at the right time, leveraging metrics such as accuracy for evaluation~\cite{li2023api, qin2023toolllm, basu2024nestful}. For example, Qin~\etal developed a benchmark for tool-augmented LLMs that assesses the model’s capability to choose the correct API and populate its parameters based on user queries~\cite{qin2023toolllm}. Although these studies often cover a wide range of domains collected in real-life, their tasks typically involve fulfilling specific requests rather than generating complete, executable software programs. 

BFCL~\cite{patil2024gorilla} is another example that provides comprehensive evaluations of LLM performance on tasks involving API calls. However, the testing scenarios in BFCL still only involve a limited set of APIs, and the API calls typically lack interdependencies between their parameters.

While these benchmarks have advanced our understanding of LLMs' function-calling abilities, they have several limitations: (1) The reliance on manually collected test cases makes updates costly and time-consuming; (2) Most benchmarks focus on simple, daily user requests rather than more complex code generation tasks that demand deeper reasoning from LLMs; and (3) Many rely on static checks, which may not fully capture the semantic correctness of the generated programs.

To address these limitations, we introduce {\ourbench}, a new benchmark built with our automated {\ourmethod} framework, which generates diverse programs directly from API documentation. Our framework enables both automatic test case generation and oracle checking, providing a more systematic way to assess LLMs' abilities to understand complex instructions and generate stateful programs. 

\subsection{Automated Testing of RESTful APIs}
Automated generation of effective and diverse API call sequences remains a commonly studied yet challenging task. 
One representative research in this field is RESTful API testing~\cite{arcuri2019restful, atlidakis2019restler, viglianisi2020resttestgen, martin2021restest, wu2022combinatorial, kim2022automated, kim2023adaptive, deng2023nautilus, golmohammadi2023testing, wei2024demystifying}, which generally follows a black-box approach and relies on standardized specifications such as OpenAPI~\cite{openapi}.
This line of work typically adopts a black-box approach, as the source code is often unavailable in API testing. In such cases, Golmohammadi~\etal identified the generation of meaningful API sequences as a primary challenge, which is complicated by inter-API dependencies and underspecified schemas~\cite{golmohammadi2023testing}.

To address this challenge, techniques for API sequence generation largely fall into two categories. Top-down and bottom-up approaches. For the former, RESTTESTGEN~\cite{viglianisi2020resttestgen} is a representative case that first constructs an overall API call graph (\eg, an Operation Dependency Graph), and then instantiates API calls based on this graph. While providing a holistic view, the efficacy of such methods can be compromised by the quality of the initial graph, which may be derived from flawed or incomplete specifications.

Conversely, bottom-up strategies~\cite{atlidakis2019restler, liu2022morest} build sequences incrementally. For example, RESTler~\cite{atlidakis2019restler} extends sequences one call at a time using heuristics, while MOREST~\cite{liu2022morest} utilizes a dynamically updating RESTful-service Property Graph to capture detailed API behaviors and dependencies. 

More recently, several studies have explored integrating LLMs into REST API test-case generation. In these approaches, LLMs are used to produce realistic test inputs, capture inter-parameter dependencies to improve test efficiency~\cite{kim2025llamaresttest, kim2025multi}, or assist in identifying valid test oracles~\cite{alonso2025satori}. While these efforts show promising results, their objectives differ fundamentally from ours: they use LLMs to enhance software testing (i.e., AI for SE), whereas our work uses software engineering techniques to evaluate LLM capabilities (i.e., SE for AI).


While RESTful API testing and our stateful testing for LLMs share some similarities in API execution trace generation, the two lines of work are orthogonal, as they address fundamentally different problems. RESTful API testing primarily focuses on data retrieval and manipulation. In contrast, {\ourmethod} is designed to evaluate LLMs’ tool use abilities, which requires modeling computations and solving related constraints. For instance, in the Tensor Manipulation scenario, {\ourmethod} must track high-dimensional tensor shapes and values in PyTorch and resolve constraints for API calls. In the text-to-speech tool usage scenario, it must manage transitions between different modalities and object properties.

\section{Conclusion}

In this paper, we introduced {\ourmethod}, an automated framework designed to evaluate the ability of large language models (LLMs) to generate code involving multiple API calls with complex interdependencies. Our approach begins with API sequence fuzzing using state machines, followed by a multi-agent process that translates code into natural language instructions. Leveraging this framework, we generated and manually curated 120 test cases across three diverse application scenarios, establishing a new benchmark, {\ourbench}.
Experimental results demonstrate that {\ourmethod} offers distinct advantages in generating sequential API calls with intricate dependencies. Evaluation of state-of-the-art LLMs on {\ourbench} reveals that there remains substantial room for improvement in their tool usage capabilities. We further identify three common root causes underlying LLM failures to generate programs that faithfully meet users' requirements. To support reproducibility and encourage further research, we will make both our codebase and the {\ourbench} benchmark publicly available. We hope that our work will serve as a foundation for future research and development, ultimately leading to LLM systems with more robust and effective tool integration.

\bibliographystyle{IEEEtran}
\bibliography{ref}

\end{document}

%% file: micro.tex
\usepackage{overpic}
\usepackage{xspace}
\usepackage{enumitem}
\usepackage{multirow}
\usepackage{mdframed}

\usepackage[utf8]{inputenc} 
\usepackage[T1]{fontenc}    
\usepackage{url}            
\usepackage{booktabs}       
\usepackage{amsfonts}       
\usepackage{nicefrac}       
\usepackage{microtype}      
\usepackage{xcolor}         
\usepackage{nameref}
\usepackage{soul}
\usepackage{color}
\usepackage{colortbl}
\usepackage{algorithm}
\usepackage{algpseudocode}
\usepackage{listings}
\usepackage{tcolorbox}
\usepackage{fancyvrb}

\newcommand{\argmax}{arg\,max}

\definecolor{darkBlue}{rgb}{0.000000,0.000000,0.545098}
\definecolor{darkGreen}{rgb}{0.000000,0.392157,0.000000}
\definecolor{DarkGray}{gray}{0.4}
\definecolor{javared}{rgb}{0.6,0,0} 
\definecolor{javagreen}{rgb}{0.25,0.5,0.35} 
\definecolor{javapurple}{rgb}{0.5,0,0.35} 
\definecolor{javadocblue}{rgb}{0.25,0.35,0.75} 
\definecolor{lightgray}{gray}{0.7}
\definecolor{lightblue}{rgb}{0.63, 0.79, 0.95}

\definecolor{shadecolor}{RGB}{150,150,150}
\definecolor{blueA}{RGB}{204,229,255}
\definecolor{redA}{RGB}{112,0, 0}
\definecolor{RED}{RGB}{255,0, 0}
\definecolor{lightred}{HTML}{FFEBEE}
\definecolor{commentgreen}{rgb}{0.0, 0.5, 0.0}

\floatstyle{plain}
\newfloat{codesnippet}{htbp}{loc}

\floatname{codesnippet}{Code Snippet}

\newfloat{example}{htbp}{loe}
\floatname{example}{Example}
\floatstyle{plain} 
\restylefloat{example}

\lstdefinestyle{mystyle}{
  frame=single,
  xleftmargin=4pt,
  xrightmargin=4pt,
  abovecaptionskip=2pt,
  belowcaptionskip=0pt,
  captionpos=b,
  escapeinside={*‘}{’*},
  tabsize=4,
  emphstyle={\bf},
  basicstyle=\linespread{0.9}\small\ttfamily,
  keywordstyle=\color{javapurple}\bfseries,
  stringstyle=\color{javared},
  commentstyle=\color{javagreen},
  morecomment=[s][\color{javadocblue}]{/**}{*/},
  showspaces=false,
  columns=flexible,
  showstringspaces=false,
  morecomment=[l]{//},
  breaklines=true
}

\lstdefinestyle{tblstyle}{
  frame=none,
  xleftmargin=-10pt,
  xrightmargin=0pt,
  abovecaptionskip=0pt,
  belowcaptionskip=0pt,
  captionpos=b,
  escapeinside={*‘}{’*},
  tabsize=4,
  emphstyle={\bf},
  basicstyle=\linespread{0.5}\fontsize{6.5}{8}\ttfamily,
  keywordstyle=\color{javapurple}\bfseries,
  stringstyle=\color{javared},
  commentstyle=\color{javagreen},
  morecomment=[s][\color{javadocblue}]{/**}{*/},
  showspaces=false,
  columns=flexible,
  showstringspaces=false,
  morecomment=[l]{//},
  breaklines=true
}

\newcommand{\code}[1]{{{\small \ttfamily #1}}}

\definecolor{lightgray}{gray}{0.7}

\definecolor{lightred}{HTML}{FFEBEE}
\definecolor{royalblue}{RGB}{65,105,225} 
\usepackage{xspace}
\makeatletter
\DeclareRobustCommand\onedot{\futurelet\@let@token\@onedot}
\def\@onedot{\ifx\@let@token.\else.\null\fi\xspace}
\def\eg{\emph{e.g}\onedot} 
\def\ie{\emph{i.e}\onedot}

\def\etal{\emph{et al}\onedot}

\makeatother

\definecolor{blue}{RGB}{0,102,204}  

\newcommand{\responseref}{}

\definecolor{lightgray}{HTML}{eeeeee}
\definecolor{darkgray}{HTML}{d8d8d8}
\definecolor{tablecellgreen}{HTML}{d0e6d0}

\definecolor{highlightColor}{rgb}{1, 0.8, 0.6}

\definecolor{amii_magenta}{HTML}{bf477c}
\definecolor{amii_summer}{HTML}{ffcccc}
\definecolor{amii_mustard}{HTML}{faa53c}
\definecolor{amii_sky}{HTML}{6c98ab}
\definecolor{amii_emerald}{HTML}{006c65}
\definecolor{amii_night}{HTML}{003f58}

\definecolor{top1Color}{HTML}{ff770f}
\definecolor{top2Color}{HTML}{01847f}
\definecolor{top3Color}{HTML}{ffecb3}
\definecolor{propColor}{HTML}{002FA7}
\definecolor{jiayang_todo}{HTML}{2FAF38}
\sethlcolor{highlightColor}

\usepackage{subcaption}

\newif\ifdisplaycontent
\displaycontenttrue  

\def\ourmethod{StateGen}
\def\ourbench{StateEval}

\definecolor{color1}{HTML}{E5E5E1}
\definecolor{color2}{HTML}{E0DDEF}
\definecolor{color3}{HTML}{F3D7CA}
\definecolor{color4}{HTML}{E7E2B6}
\definecolor{color5}{HTML}{EAD6FA}

\newcommand{\circled}[1]{{\large \textcircled{\footnotesize #1}}}

\newcounter{finding}
\newenvironment{finding}
{
    \refstepcounter{finding}
	\begin{mdframed}[
    	nobreak=true,
    	linecolor=black,
    	roundcorner=12pt,
    	backgroundcolor=gray!05,
    	linewidth=0.5pt,
    	leftmargin=1pt,
    	rightmargin=1pt,
        innerleftmargin=5pt,
        innerrightmargin=5pt,
    	topline=true,
    	bottomline=true,
    	skipabove=10pt
	]
    \textbf{Answer to RQ\arabic{finding}}: 
}
{
    \end{mdframed}
    \vspace{3pt}
}

%% file: algorithm.tex
\begin{algorithm}[t]
\caption{Trace Generation with Coverage Guidance}
\label{alg:trace_gen}
\begin{algorithmic}[1]
\Require $n \in \mathbb{N}^+$ \Comment{Target API call count}
\Ensure $O$ \Comment{Generated trace of API calls}

\Procedure{GenerateTrace}{$n$}
    \State $O \gets \emptyset$, $S \gets RandomInit$ \Comment{Initialization}
    \State $\hat{o} \gets \bot$ \Comment{Initialize previous transition to None}
    
    \For{$i \gets 1$ \textbf{to} $n$}
        \State $\mathcal{A} \gets \{o \in \mathcal{O} \mid \text{valid}(o, S, O)\}$ \Comment{Obtain transitions}
        \State $\mathcal{E} \gets \left\{\frac{1}{\nu((\hat{o}, o))+\epsilon} \mid o \in \mathcal{A}\right\}$ \Comment{Energy weights}
        \State $\mathcal{C} \gets \{o \in \mathcal{A} \mid \nu((\hat{o}, o))=0\}$ \Comment{New coverage}
        
        \If{$\mathcal{C} \neq \emptyset$}
            \State $o^* \gets \argmax_{o  \in \mathcal{C}} \phi(o,S)$  \Comment{$\phi$ indicates pair transition coverage}
        \Else
            \State $o^* \sim P(o) \propto \mathcal{E}$ \Comment{Energy-based sampling}
        \EndIf
        
        \State $\nu \gets \nu((\hat{o}, o^*)) + 1$ \Comment{Update frequency recorder}
        \State $S \gets \delta(S, o^*)$ \Comment{Update state schema}
        \State $O \gets O \circ o^{*}$ \Comment{Append to trace}
        \State $\hat{o} \gets o^*$
    \EndFor
    
    \State \Return $O$
\EndProcedure
\end{algorithmic}

\end{algorithm}

%% file: table/dataset_statistics.tex
\begin{table*}[htbp]
    \centering
    \footnotesize
    \responseref{}
    \setlength{\tabcolsep}{4pt}
    \begin{tabular}{llccccc}
    \toprule
        \multicolumn{2}{c}{\textbf{Statistics}} & HumanEval & DS-1000  & BigCodeBench & BFCL & \textit{\ourbench}  \\

        \midrule
        
        \multicolumn{2}{c}{\textbf{Instruction Length}} & 131.32 & 282.42 & 144.83 & 77.07 & 717.08 \\

        \multicolumn{2}{c}{\textbf{Code Length}} & 53.85 & 42.06 & 112.44 & 44.30 & 442.11 \\

        \multicolumn{2}{c}{\textbf{Function Call Num}} & 0.18 & 2.07 & 3.68 & 2.50 & 7.16 \\

        \multicolumn{2}{c}{\textbf{Path Depth}} & 0.12 & 0.37 & 0.48 & 0.00 & 2.00 \\

        \multicolumn{2}{c}{\textbf{Binding Count}} & 0.05 & 1.17 & 1.07 & 0.00 & 5.03 \\
    
    \bottomrule
    \end{tabular}
    \caption{
    Statistics averaged at the sample level for different benchmarks. We use \textit{cl100k\_base} tokenizer from OpenAI to count the token length for the instruction and reference code. 
    }
    
    \vspace{-5mm}

    \label{tab:dataset}
\end{table*}